\documentclass[a4paper,onecolumn,11pt,tightenlines,superscriptaddress,showkeys,showpacs]{revtex4-1}

\usepackage{epsfig}
\usepackage{graphicx}
\usepackage{pslatex}
\usepackage{color}
\usepackage{amssymb}
\usepackage{xspace}
\usepackage{amsmath}

\newcommand{\red}{\textcolor{black}}

\newcommand*{\sqs}{\ensuremath{\sqrt{s}}\xspace}
\newcommand*{\Nch}{\ensuremath{N_\mathrm{ch}}\xspace}
\newcommand*{\pT}{\ensuremath{p_\mathrm{T}}\xspace}
\newcommand*{\kT}{\ensuremath{k_\mathrm{T}}\xspace}
\newcommand*{\pTjet}{\ensuremath{p_\mathrm{T}^{\mathrm{jet}}}\xspace}
\newcommand*{\GeV}{\ensuremath{\mathrm{GeV}}\xspace}
\newcommand*{\TeV}{\ensuremath{\mathrm{TeV}}\xspace}

\newcommand*{\rhoMB}{\ensuremath{\rho_\mathrm{MI}}\xspace}

\begin{document}

\title{Modification of jet structure in high-multiplicity pp collisions due to multiple-parton interactions and observing a multiplicity-independent characteristic jet size}

\author{Zolt\'{a}n Varga}\affiliation{Wigner Research Centre for Physics, P.O Box 49, H-1525 Budapest, Hungary}\affiliation{Department of Theoretical Physics, Budapest University of Technology and Economics, 8 Budafoki Rd., H-1521 Budapest, Hungary}
\author{R\'{o}bert V\'ertesi}\email[Correspondence should be addressed to R\'{o}bert V\'ertesi: ]{vertesi.robert@wigner.mta.hu}\affiliation{Wigner Research Centre for Physics, P.O Box 49, H-1525 Budapest, Hungary}
\author{Gergely G\'{a}bor Barnaf\"{o}ldi}\affiliation{Wigner Research Centre for Physics, P.O Box 49, H-1525 Budapest, Hungary}

\date{\today}

\begin{abstract}
We study the multiplicity dependence of jet structures in pp collisions using Monte Carlo event generators. We give predictions for multiplicity-differential jet structures and present evidence for a non-trivial jet shape dependence on charged hadron event multiplicity, that can be used as a sensitive tool to experimentally differentiate between equally well-preforming simulation tunes. We also propose a way to validate the presence and extent of effects such as multiple parton interactions (MPI) or color reconnection (CR), based on the detection of non-trivial jet shape modification in high-multiplicity events at high \pT. Using multiplicity-dependent jet structure observables in various \pT windows might also help understanding the interplay between jet particles and the underlying event (UE). \red{We introduce a multiplicity-independent characteristic jet size measure, and use a simplistic model to aid its physical interpretation.}
\end{abstract}

\maketitle

\section{Introduction}

High multiplicity events of small colliding systems at high center-of-mass energies show similar collective features to those observed in events of heavy ion collisions with comparable multiplicities, such as long-range near-side correlations and $v_n$ ("flow") coefficients~\cite{Yan:2013laa,Khachatryan:2010gv}.
Whether this behavior may be attributed to the presence of a deconfined state in small systems is an open question. However, possible medium-like effects in pp may question the widely exploited assumption that pp collisions are safe to use as a reference for heavy-ion systems. On the other hand, recent studies showed that flow patterns may emerge from features different than hydrodynamics. For instance, radial flow in pp collisions may be explained by pure QCD mechanisms such as multiple-parton interactions (MPI)~\cite{Ortiz:2016kpz}. Such a case questions signatures previously considered as definite signs of the QGP.
Recent analyses of pp and p-Pb collisions also show a universal enhancement of heavy-flavour particles, that is usually attributed to MPI and higher gluon radiation associated with short distance production processes~\cite{Adam:2016mkz}. However, we lack the qualitative understanding of these effects.
While we cannot expect to observe direct modification of particle yields by any medium created in collisions of small systems (because of the small volume of such a medium), phenomena that act in the soft-hard transitional regime should in principle pose an effect on hard processes as well. A modification in the shapes of developing jets that can in principle be accessible by existing experiments.

Jet profile measurements in hadron colliders have long been suggested as sensitive probes of QCD parton splitting and showering calculations~\cite{Seymour:1997kj,Ellis:1992qq,Vitev:2008rz}, and even as an indicator of the QGP~\cite{Vitev:2008rz}. 
\red{A recent study suggests to verify a possible existence of a QGP-droplet by measuring properties of jets in association with a $Z$-boson in ultra-central pp collisions~\cite{Mangano:2017plv}.}
In experiment, jet structure observables with full jet reconstruction have been studied in different collisional systems at HERA, the Tevatron and the LHC~\cite{Adloff:1998ni,Breitweg:1997gg,Abe:1992wv,Abachi:1995zw,Acosta:2005ix,Aad:2011kq,Chatrchyan:2012mec,Chatrchyan:2013ala} among others.
\red{It is especially important to gain a detailed understanding on multiplicity-dependence of the jet structures up to high momenta with the recent advent of machine learning classification techniques in jet studies~\cite{Guest:2016iqz,Haake:2017dpr}, since these rely heavily on the modelling of parton shower and fragmentation and their connection to the underlying event, in order to avoid possible selection biases.}

We use the PYTHIA event generator~\cite{Sjostrand:2014zea} to extensively study the multiplicity-dependent jet shapes, using different tunes and setups of PYTHIA to examine the possible effects of MPI on jets. We provide predictions for pp collisions at $\sqs=7$~\TeV to motivate similar, multiplicity-dependent jet-structure measurements at the LHC. In models with string hadronization, the recombination of overlapping color strings (color reconnection or CR) influence fragmentation and are also known to produce collective-like patterns such as radial flow~\cite{Bierlich:2014xba}. We investigate the effects caused by the choice of the CR scheme within PYTHIA on the simulated jet structures. As a reference point in our investigations, we decided to use a set of jet structure measurements by the CMS experiment at $\sqs=7\ \TeV$, carried out in a wide jet momentum range from 15~GeV/$c$ up to 1000~GeV/$c$~\cite{Chatrchyan:2012mec}. A previous CMS study has investigated multiplicity-differential jet structures, albeit momentum-inclusively with a $\pTjet>5$ GeV/$c$ jet transverse momentum threshold, to understand the influence of underlying events (UE) on jets~\cite{Chatrchyan:2013ala}.

Our paper is organized as follows. In Section~\ref{sec:simulation}.\ we describe our analysis method in details, and show its validation on CMS data. In Section~\ref{sec:results}.\ we present and discuss our results, complemented by simplistic model calculations that aid the understanding of the physics implications. Finally we summarize our results in Section~\ref{sec:summary}.

\section{Simulation and Analysis}
\label{sec:simulation}

We used the PYTHIA 8.226~\cite{Sjostrand:2014zea} event generator to generate random pp collisions at a center-of-mass energy of $\sqs=7\ \TeV$. We allowed any hard pQCD process, but in order to decrease simulation time we limited the phase space by requiring a certain minimum invariant transverse momentum $\hat{\pT}$ of the hardest $2\rightarrow 2$ process in an event. We chose $\hat{\pT}>5$ GeV/$c$,  $\hat{\pT}>20$ GeV/$c$, $\hat{\pT}>40$ GeV/$c$ and $\hat{\pT}>80$ GeV/$c$ for the evaluation of jets with $\pTjet>15$ GeV/$c$, $\pTjet>50$ GeV/$c$, $\pTjet>80$ GeV/$c$ and $\pTjet>125$ GeV/$c$, respectively. These cutoffs were determined so that they do not have influence on the shape of the reconstructed \pTjet spectrum. We simulated 5 million events with each of the settings.

Since many physical details can not be derived from first principles due to our limited understanding of Nature, the MC event generators, including PYTHIA, require extra input parameters. Determining these parameters are far from trivial, and a given set of the parameters are generally sufficient only for reproducing certain experimental data. A given configuration of these parameters, optimized for reproducing experimental results in certain physical aspects, are called tunes. Besides the default tune Monash 2013 (Monash) we also investigated two others, the Monash* and 4C tunes. The Monash tune, which uses the NNPDF2.3LO PDF set~\cite{Ball:2013hta}, is specifically configured to both $e^+e^-$ and pp/p$\bar{\textrm{p}}$ data~\cite{Skands:2014pea}. Monash* (or CUETP8M1-NNPDF2.3LO) is an underlying-event tune based on the Monash tune and was configured to CMS data~\cite{Khachatryan:2015pea}. The 4C tune is a newer one introduced with PYTHIA version 8.145~\cite{Corke:2010yf}. It is based on the tune 2C, but it uses the CTEQ6L1 PDF set~\cite{Pumplin:2002vw} and has further changes including a reduced cross section for diffraction and modified multi-parton interaction parameters to produce a higher and more rapidly increasing charged pseudorapidity plateau for better agreement with some early key LHC numbers~\cite{Buckley:2011ms}.
Using the Monash tune as a starting point we also did investigations where we changed some settings in PYTHIA to directly study their effect on the jet structure. There are continuously developed models of multiple-parton interactions implemented in PYTHIA~\cite{Sjostrand:1987su,Sjostrand:2017cdm}. To understand the multiplicity-dependent jet modification by MPI we used data samples where we switched this effect on and off.

We also investigated different color reconnection schemes provided by PYTHIA, including turning off this feature. Color reconnection is a built-in mechanism in PYTHIA that allows interactions between partons originating in MPI and initial/final state radiations, by minimizing color string length. Since this procedure is quite ambiguous, several models are implemented. The original MPI-based scheme used in PYTHIA 8.226 (that we denote CR0 in the followings) relies on the parton shower-like configuration of the beam remnant. In an additional step, it merges the gluons of a lower-$p_T$ MPI system with gluons of a higher-$p_T$ MPI system. A newer QCD-based scheme~\cite{Christiansen:2015yqa} (CR1) relies, however, on the full QCD color configuration in the beam remnant. Then the color reconnection is made by minimizing the potential string energy. The QCD color rules are incorporated in the CR to determine the probability that a reconnection is allowed. This model also allows the creation of junction structures. Besides the above mentioned CR schemes, a so-called gluon move scheme~\cite{Argyropoulos:2014zoa} (CR2) has been implemented to PYTHIA recently, in which gluons can be moved from one location to another so as to reduce the total string length.

We carried out a full jet reconstruction including both charged and neutral particles, using three popular algorithms, the anti-\kT{}~\cite{Cacciari:2008gp}, \kT{}~\cite{Catani:1993hr,Ellis:1993tq} and Cambridge-Aachen~\cite{Dokshitzer:1997in,Wobisch:1998wt} algorithms, provided by the FASTJET~\cite{Cacciari:2011ma} software package. All of them are sequential clustering algorithms, meaning that the closest particle tracks in momentum space are sequentially merged one-by-one according to the minimum of a distance measure between the particle four-momentums.
While all three algorithms are infrared- and collinear-safe, in high-multiplicity environments the clusterization outcomes will be rather different. Anti-\kT is popular because it is only slightly susceptible to pile-up and underlying events, and it clusterizes hard jets into nearly perfect cones with a resolution parameter $R$ even in high-multiplicity events, in accordance with the general image of how a jet should look like. The other two algorithms are more suitable for jet substructure studies but provide jets of irregular shape that are not uniform in area, especially the \kT algorithm, where the area of the jets fluctuates considerably~\cite{Cacciari:2011ma}. Similarly to the CMS analysis~\cite{Chatrchyan:2012mec,Chatrchyan:2013ala}, we selected inclusive jets, with a resolution parameter $R=0.7$. We considered constituent particles, with a transverse momentum threshold $|p_{\mathrm{T},\mathrm{track}}| > 0.15\ \GeV/c$, at the generator level. Our experience matches earlier findings that the detector effects, after corrections, do not change the simulated jet observables significantly~\cite{Chatrchyan:2012mec}.
We examined jets in the pseudorapidity window $|\eta_\mathrm{jet}|<1$. We restricted our investigations to the $15\ \GeV/c<\pTjet<400\ \GeV/c$ jet momentum range, where multiplicity-differential studies are feasible in the near future. 

For the investigation of a possible jet shape modification we analyze the transverse momentum profile of the jets. Two widely used observables are the differential jet shape ($\rho$) and the integral jet shape ($\psi$). The differential jet shape describes the radial transverse momentum distribution inside the jet cone and is defined as follows:
\begin{equation}
\rho(r) = \frac{1}{\delta r} \frac{1}{\pTjet}\sum\limits_{\substack{r_a < r_i < r_b}} p_{\mathrm{T}}^i,
\end{equation}
where $p_{\mathrm{T}}^i$ is the transverse momentum of a particle inside a $\delta r$ wide annulus with inner radius $r_a = r - \frac{\delta r}{2}$ and outer radius $r_b = r + \frac{\delta r}{2}$ around the jet axis and \pTjet is the transverse momentum of the whole jet. The distance of a given particle from the jet axis is given by $r_i = \sqrt{(\phi_i - \phi_\mathrm{jet})^2 + (\eta_i - \eta_\mathrm{jet})^2}$, where $\phi$ is the azimuthal angle and $\eta$ is the pseudorapidity. The integral jet shape gives the average fraction of the jet transverse momentum contained inside a cone of radius $r$ around the jet axis and is calculated as
\begin{equation}
\psi(r) = \frac{1}{ \pTjet }\sum\limits_{\substack{r_i < r}} p_{\mathrm{T}}^i,
\end{equation}
where the symbols denote the same quantities as for the differential jet shape.

\begin{figure}[!h]
\includegraphics[width=0.33\linewidth]{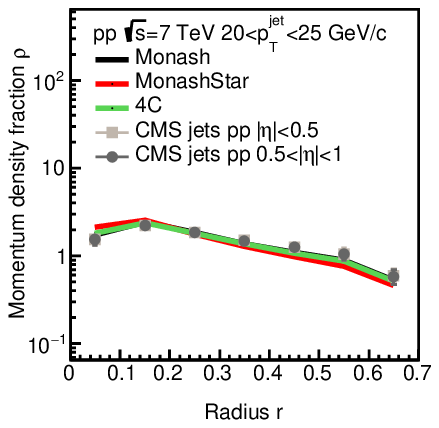}%
\includegraphics[width=0.33\linewidth]{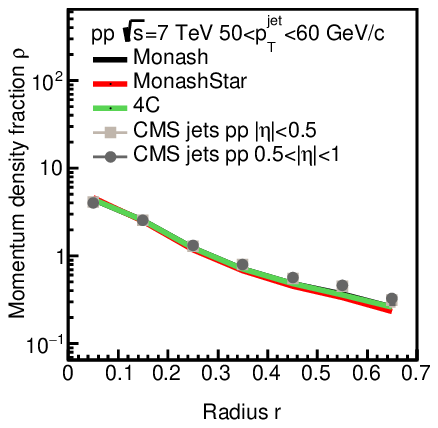}%
\includegraphics[width=0.33\linewidth]{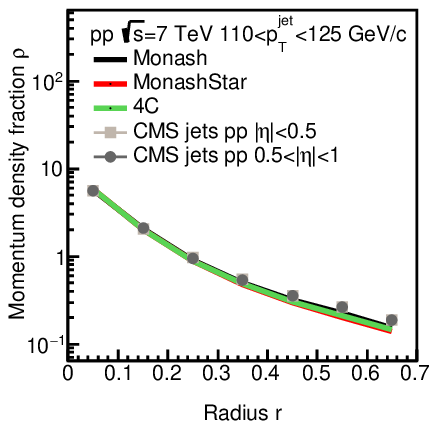}%
\caption{\label{rhoCMS} Differential jet structure $\rho(r)$ measured by the CMS experiment in pp collisions at \sqs=7 TeV~\cite{Chatrchyan:2012mec}, compared to different PYTHIA tunes, for $20\ \GeV/c <\pTjet<25$ \GeV/$c$ {\it (left)}, $50\ \GeV/c<\pTjet<60\ \GeV/c$ {\it (center)} and $110\ \GeV/c<\pTjet<125\ \GeV/c$ {\it (right)}.}
\end{figure}

As a first step we showed that our simulations reproduce CMS data~\cite{Chatrchyan:2012mec} within uncertainty throughout this range. We show examples in three different \pTjet windows in Fig.~\ref{rhoCMS}.
For harder jets, the calculated momentum density distribution gets steeper in the central (small-$r$) region of the jets, in qualitative accordance with the calculations of Ref.~\cite{Vitev:2008rz}.

\begin{figure}[!h]
\includegraphics[width=0.33\linewidth]{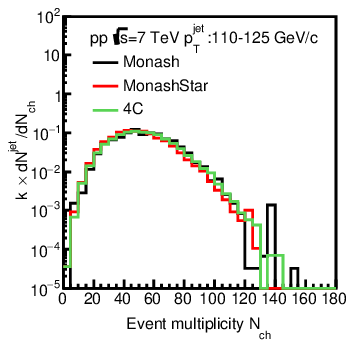}%
\includegraphics[width=0.33\linewidth]{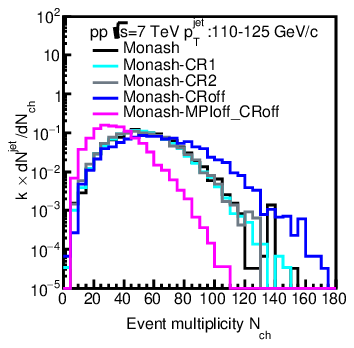}
\caption{\label{mults} Distributions of event multiplicity for jets in the $110\ \GeV/c<\pTjet<125 \ \GeV/c$ window, compared for the Monash, Monash* and 4C tunes {\it (left)}; and for the Monash tune with the CR0, CR1, CR2, settings as well as CR turned off and MPI turned off {\it (right)}.}
\end{figure}

\begin{figure}[!h]
\includegraphics[width=0.33\linewidth]{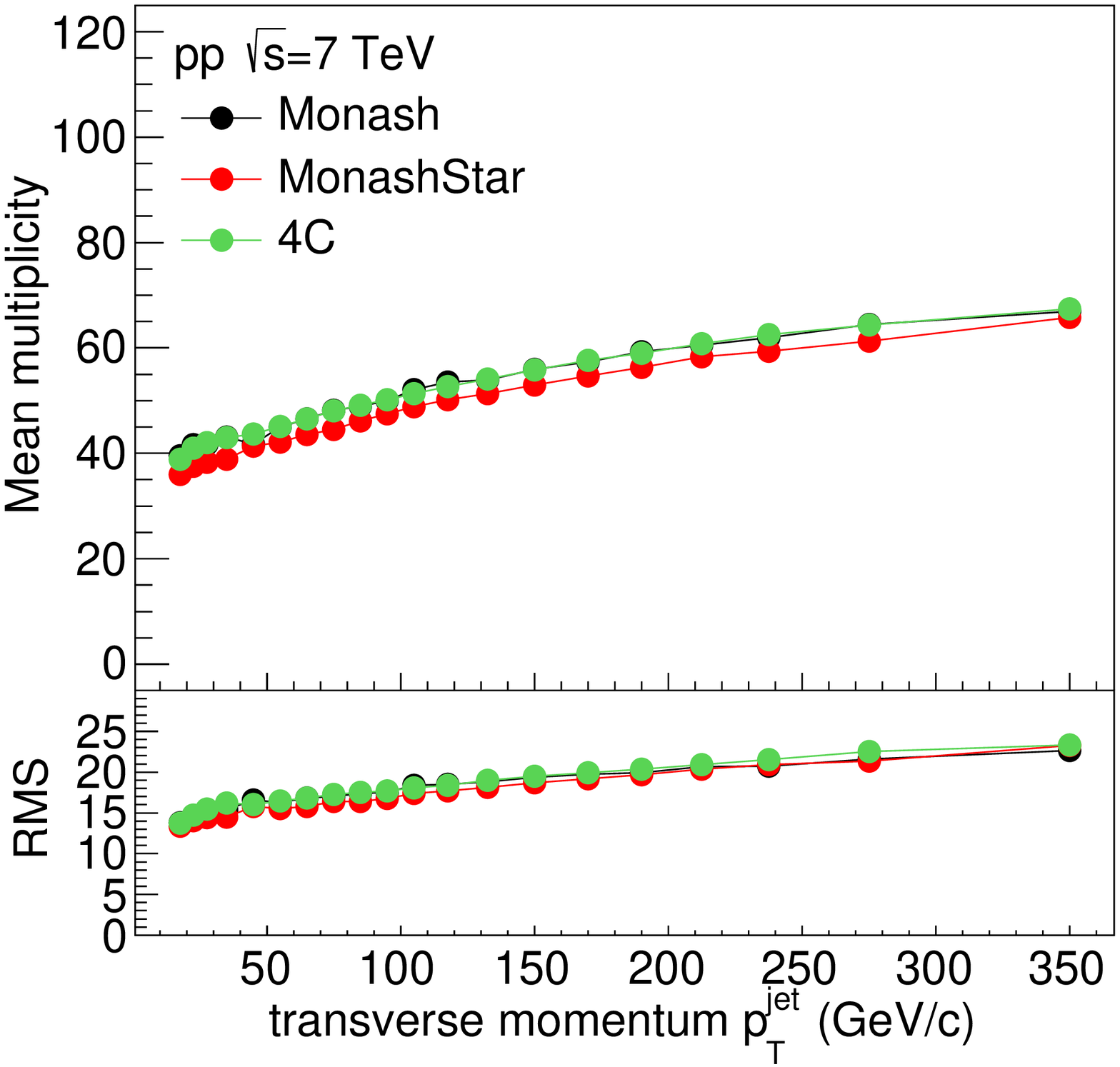}%
\includegraphics[width=0.33\linewidth]{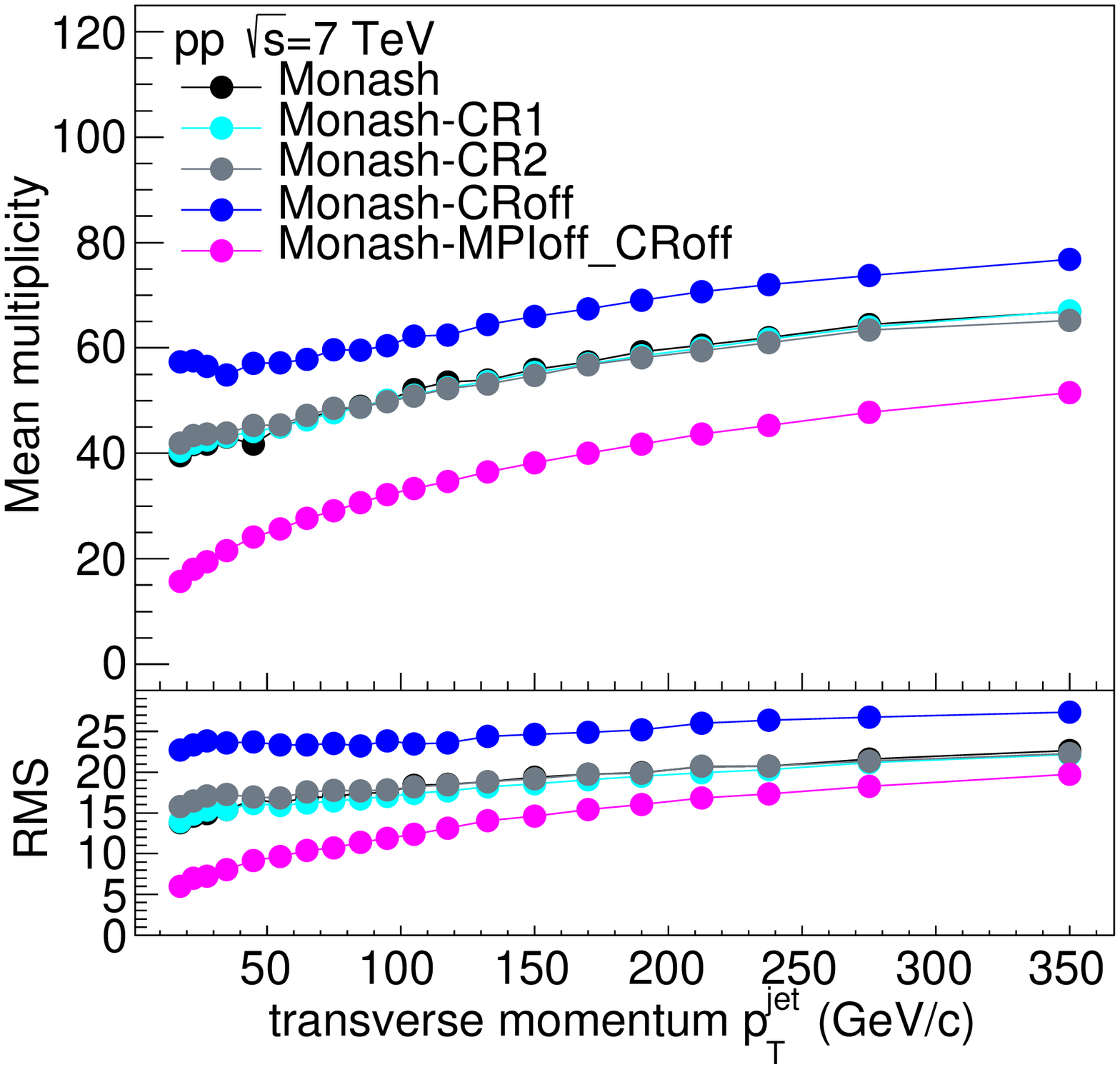}%
\caption{\label{multmeansrms} Mean and RMS values of the event multiplicity distributions for jets depending on \pTjet, compared for different tunes {\it (left)} and settings {\it (right)}. The uncertainities of the mean and RMS values are smaller than the symbol size.}
\end{figure}

We investigate the jet structure for different charged hadron multiplicity (\Nch) classes. 
Generally, PYTHIA is known to reproduce multiplicities in LHC data with little differences over a broad \pT range. The CUETP8M1 and Z2* tunes reproduce pion and kaon average \pT versus track multiplicities within errors~\cite{Sirunyan:2017zmn}. The D6T and Z2 tunes show a marginal agreement with the CMS jet-multiplicity data, with about 5\% higher predictions than the mean values, flat in \pT~\cite{Chatrchyan:2012mec}.
We use charged hadron multiplicity at mid-rapidity (referred to as multiplicity in the followings for the sake of simplicity), defined as the number of the charged final state particles with $|\eta|<1$ in a given event. We show the multiplicity distributions in Fig.~\ref{mults} for the jet momentum window $110\ \GeV/c <\pTjet<125\ \GeV/c$ as an example. As shown in the left panel, distributions of the multiplicity are very similar for the different tunes. However, when considering the multiplicity distribution from different settings of the Monash tune, shown on the right panel, a substantial difference can be seen between the settings with and without MPI or CR. Disabling MPI (and CR, which assumes MPI) causes the distribution to shift towards lower values, while keeping a similar shape. Disabling CR only, on the other hand, causes the multiplicity distribution to extend toward higher values.
This means that care should be taken when one compares distributions with MPI or CR settings on and off, as it may be biased when the chosen multiplicity class is too wide. We note that multiplicity distributions from different color reconnection schemes do not differ significantly.
The \pTjet dependence of the mean and RMS values of the multiplicity distribution is compared in Fig.~\ref{multmeansrms} for different tunes, as well as for different settings in the case of the Monash tune. 
The three tunes predict very similar mean and RMS values throughout the \pTjet range. While the means of the 4C and Monash tunes overlap, Monash* predicts slightly lower multiplicities. Again, MPI and CR have a grave effect on the distributions. Swiching off MPI causes a downward shift of about 15 to 25 in mean \Nch at any \pTjet, or almost a factor of three at low \pTjet values, while switching off CR alone causes a somewhat less drastic increase of about 10 to 20 in mean \Nch counts.
 
\section{Results}
\label{sec:results}

\begin{figure}[!h]
\includegraphics[width=0.33\linewidth]{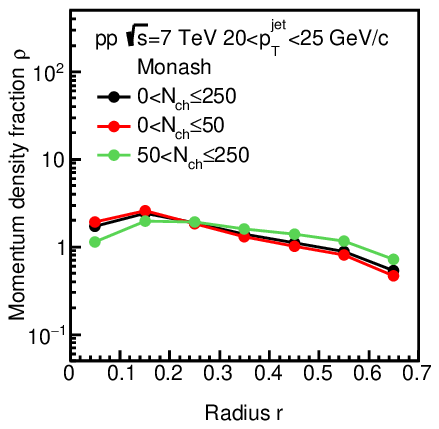}%
\includegraphics[width=0.33\linewidth]{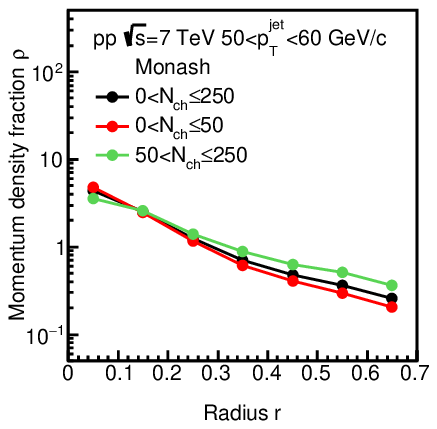}%
\includegraphics[width=0.33\linewidth]{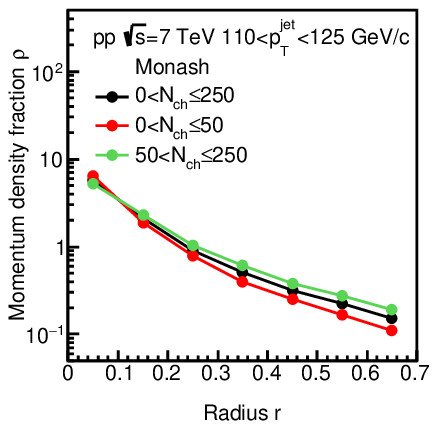}%
\caption{\label{rhomult}
Simulated differential jet structure $\rho(r)$ compared in multiplicity-integrated events ({\it black}), low-multiplicity ($\Nch \leq 50$, {\it red}) and high-multiplicity events ($\Nch>50$, {\it green}), for $20\ \GeV/c<\pTjet<25\ \GeV/c$ {\it (left)}, $50\ \GeV/c <\pTjet<60\ \GeV/c$ {\it (center)} and $110\ \GeV/c<\pTjet<125\ \GeV/c$ {\it (right)}.}
\end{figure}

In this section we present our results and consider the possible physical implications. As a first step we compute $\rho(r)$ similarly to Fig.~\ref{rhoCMS}, but this time while dividing up the data into two multiplicity classes, $\Nch\le 50$ and $\Nch >50$, respectively. We see a multiplicity dependence in the jet shapes in Fig. \ref{rhomult}. Namely, the jets contain a higher fraction 
of their transverse momentum closer to their axis and a lower fraction 
further away from their axis in the case of low multiplicity. For high multiplicity the jet shape behaves in the opposite way. This is a trivial, expected multiplicity dependence arising from two reasons. The first one is that event multiplicity is correlated with jet multiplicity, resulting in a higher fraction of narrow jets in low-\Nch events. The second reason is the UE background, which affects the jet structure \red{more} at higher $r$ values, and its effect is stronger in the case of high-\Nch events.

\begin{figure}[!h]
\includegraphics[width=0.33\linewidth]{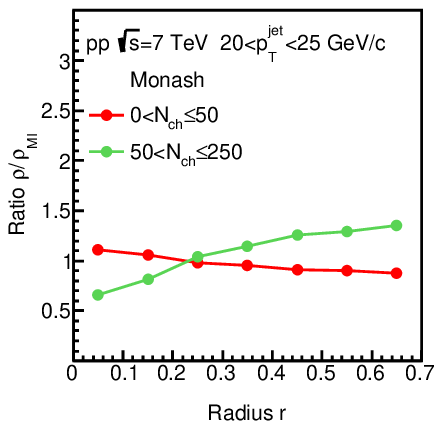}%
\includegraphics[width=0.33\linewidth]{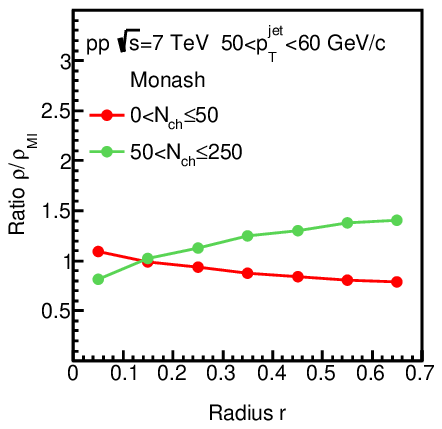}%
\includegraphics[width=0.33\linewidth]{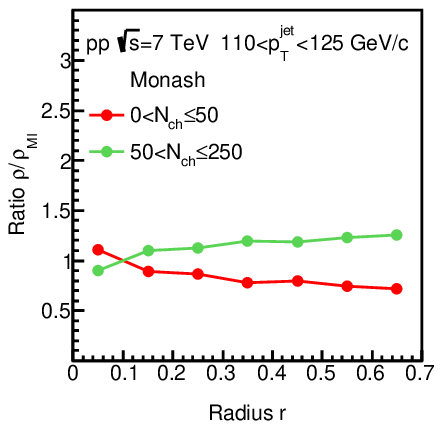}%
\caption{\label{rhoratio1} 
Ratio $\rho/\rhoMB$ of differential jet structure in low-multiplicity ($\Nch \leq 50$, {\it red}) and high-multiplicity events ($\Nch>50$, {\it green}) over multiplicity-integrated events, for $20\ \GeV/c <\pTjet<25\ \GeV/c$ {\it (left)}, $50\ \GeV/c <\pTjet<60\ \GeV/c$ {\it (center)} and $110\ \GeV/c<\pTjet<125\ \GeV/c$ {\it (right)}.}
\end{figure}

Measurements by the CMS experiment~\cite{Chatrchyan:2013ala}, that compare five multiplicity classes within the range $10<\Nch\leq 140$ and reconstruct jets at momenta $\pTjet > 5\ \GeV/c$, saw a remarkable difference between low and high-multiplicity $\rho(r)$ at low $r$ values. We can make the same observation at relatively low \pTjet values (Fig.~\ref{rhomult} left panel). Dividing $\rho(r)$ for both the high- and low-multiplicity classes with the multiplicity-integrated $\rhoMB(r)$ (no condition on \Nch), shown in Fig.~\ref{rhoratio1}, highlights this trend. The curves are much more apart at small \pTjet for low $r$ values, while there is relatively little difference between different \pTjet windows at high $r$. This suggests that jets in high multiplicity events contain much more contribution from the soft regime, and soft physics is selected by a lower choice of momentum range.

\begin{figure}[!h]
\includegraphics[width=0.33\linewidth]{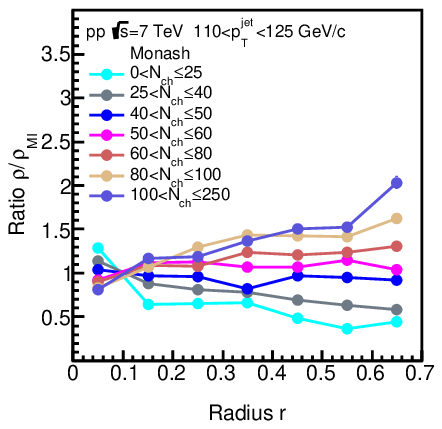}%
\includegraphics[width=0.33\linewidth]{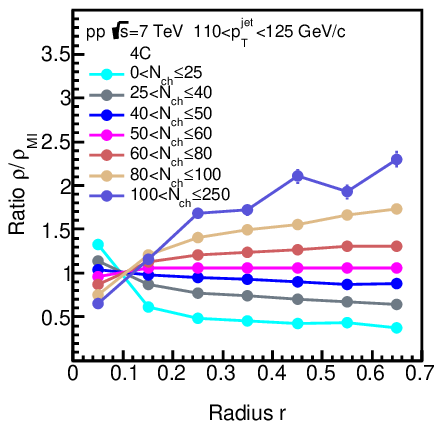}%
\includegraphics[width=0.33\linewidth]{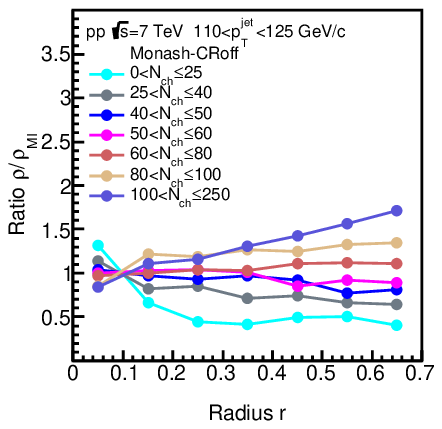}%
\caption{\label{rhoratio2} 
Ratio $\rho/\red{\rhoMB}$ of differential jet structure in several multiplicity classes (see legend) over multiplicity-integrated events, for jets within the $110\ \GeV/c<\pTjet<125\ \GeV/c$ window. The {\it left} hand side, {\it central} and {\it right} panels show events generated using the Monash tune, the 4C tune, and the Monash tune without CR respectively.}
\end{figure}

Jets in low-multiplicity events are on average narrower than in high-multiplicity events, hence the corresponding $\rho(r)/\rhoMB(r)$ ratio is above unity, while for high-multiplicity events this ratio is below unity. At high $r$ values, where UE tracks give a non-negligible contribution especially in the high-multiplicity events, the situation is just the opposite. In between there is a point at a given $r$ value where the two curves intersect each other at unity, meaning that at that radius the jets are just average. 
In Fig.~\ref{rhoratio1} we see three examples in different \pTjet windows and we can observe that the intersection point is dependent on the jet momentum. 
This is not unexpected since harder jets are narrower and UE is significant already at smaller radii.
To have a closer look at this behavior we evaluate $\rho(r)/\rhoMB(r)$ in a more refined division of data with seven multiplicity classes in the range $1\le\Nch\le 250$. We find that the curves intersect unity at virtually the same location for a given \pTjet value. This statement holds even if we compare different PYTHIA tunes and MPI or CR settings, as shown on the examples in Fig.~\ref{rhoratio2} for the Monash and 4C tunes as well as the Monash tune without color reconnection. 

\begin{figure}[!h]
\includegraphics[width=0.33\linewidth]{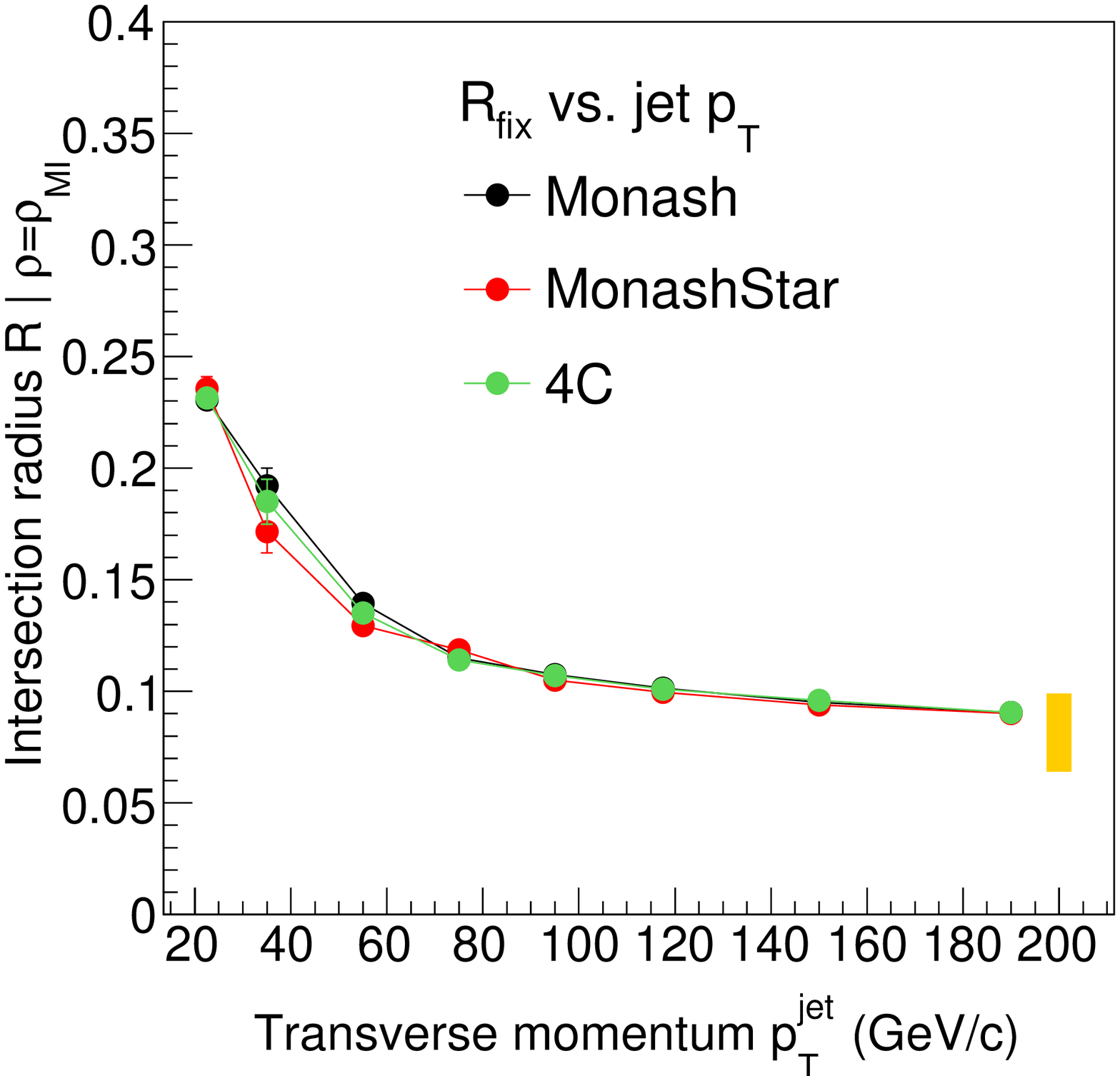}%
\includegraphics[width=0.33\linewidth]{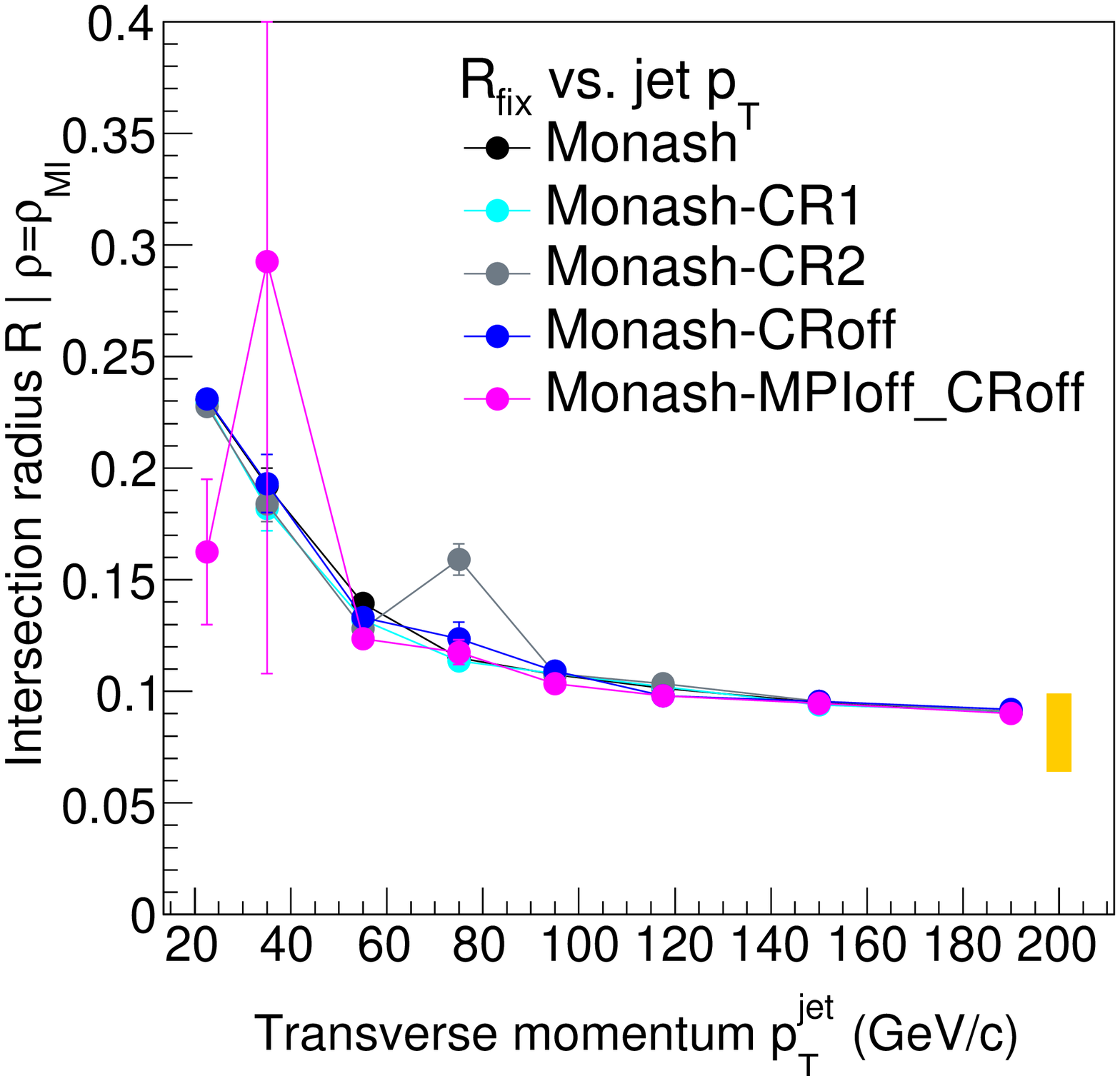}%
\includegraphics[width=0.33\linewidth]{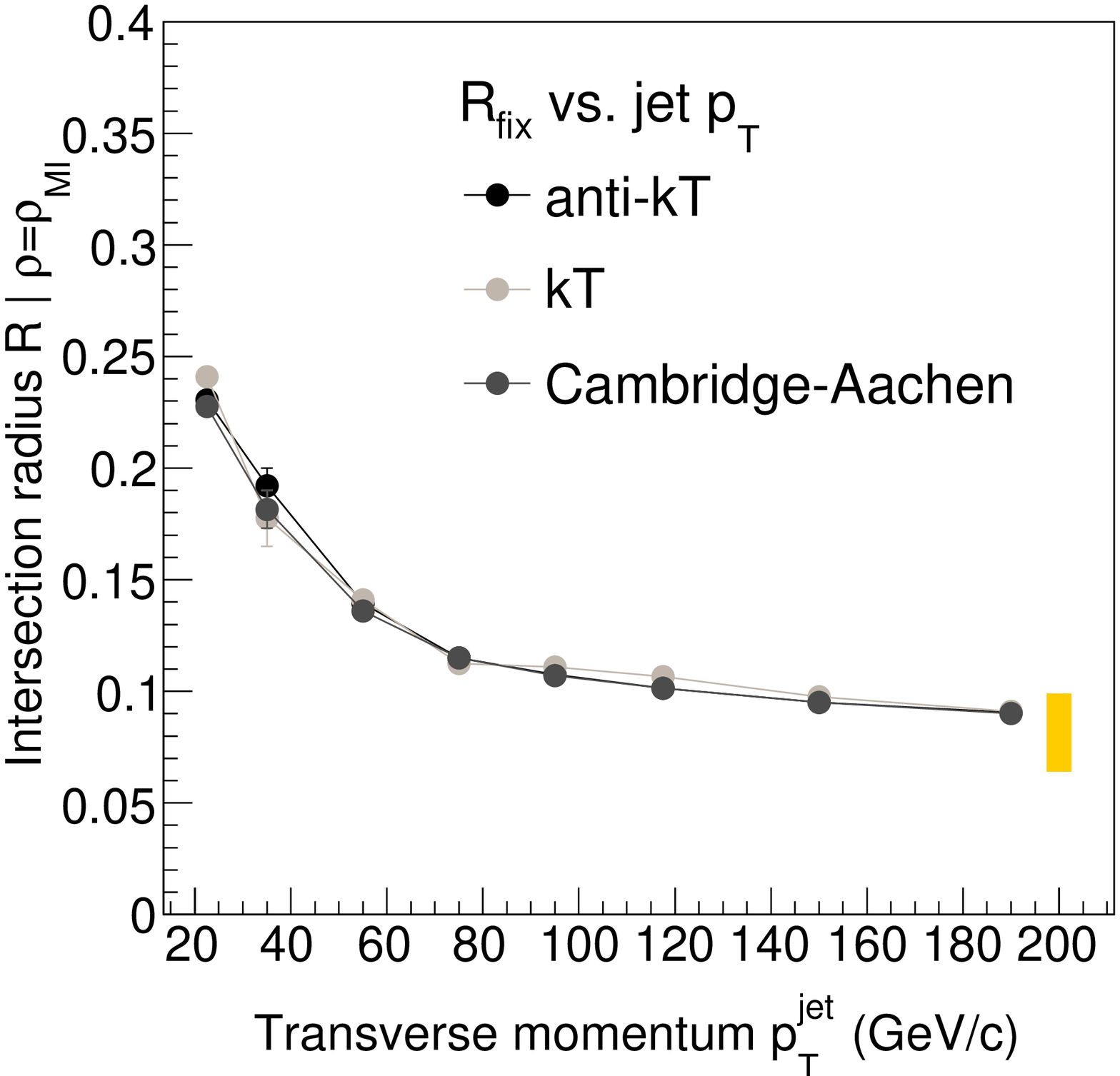}%
\caption{\label{interpolateratio}
Evolution of the intersection radius $R_\mathrm{fix}$ with the jet momentum \pTjet, compared for several PYTHIA tunes {\it (left)}, different settings {\it (center)}, and for different jet reconstruction algorithms in the case of the Monash tune {\it (right)}. The range indicated with the shaded band is the absolute uncertainty arising from the choice of bin width. (See text for details).}
\end{figure}

In the left and center panels of Fig.~\ref{interpolateratio} we plot the $r$ dependence of the intersection radius w.r.t. the jet transverse momentum for different tunes, as well as the different settings of the Monash tune. The intersection radius is computed using a linear interpolation between the two nearest points of $\rho(r)/\rhoMB(r)$, and its uncertainty is estimated by taking both the high and the low-multiplicity classes, moving the points to the upper and lower edge of their error bars in both cases, and determining the maximal and minimal values of the intersection  radius from these cases. We observe that for all tunes and settings that we tested, the intersection radii are consistent within uncertainties for any chosen \pTjet value.
There is additional uncertainty on the obtained intersection radius stemming from the linear interpolation between finite, $\delta r = 0.1$ wide bins. In order to estimate this, we repeated the analysis with the three tunes in $\delta r = 0.05$ wide bins. While the statistical fluctuations increase, the points move a maximum of 4\% upwards or 28\% downwards in a strongly correlated manner (see Fig.~\ref{interpolateratio}). Nevertheless, the overall shape of the curves remain very similar, and statistically consistent between different tunes point-by-point.
Therefore, we suggest that the intersection radius $R_\mathrm{fix}=r|_{\rho=\rhoMB}$ be considered as a characteristic jet size measure specific for a given jet transverse momentum. \red{We note that the value of $R_\mathrm{fix}$ should not be compared to the resolution parameter $R$, that is typically chosen so that most of the jet momentum is contained within the radius $R$. In contrast, $R_\mathrm{fix}$ is defined as a radius where the momentum density of the jet from events of any multiplicity is just like in the average jet, and substantial fraction of jet momentum falls towards smaller as well as towards larger radii.}

Jet shapes depend on the jet reconstruction algorithm, so we investigated whether the observed stability of the intersection radius can be an artifact of the jet reconstruction algorithm itself. Besides the anti-\kT algorithm we first used, we have reprocessed all the data with using the \kT and the Cambridge-Aachen algorithms. 
We do not find a significant difference beyond the statistics-driven fluctuations between data reconstructed by different clusterization algorithms in any of the tunes or MPI/CR settings. In the right panel of Fig.~\ref{interpolateratio} we show a comparison of $R_\mathrm{fix}(\pTjet)$ for the Monash tune with the three different algorithms.
 
\begin{figure}[!t]
\hspace{0.06\linewidth}
\includegraphics[width=0.19\linewidth]{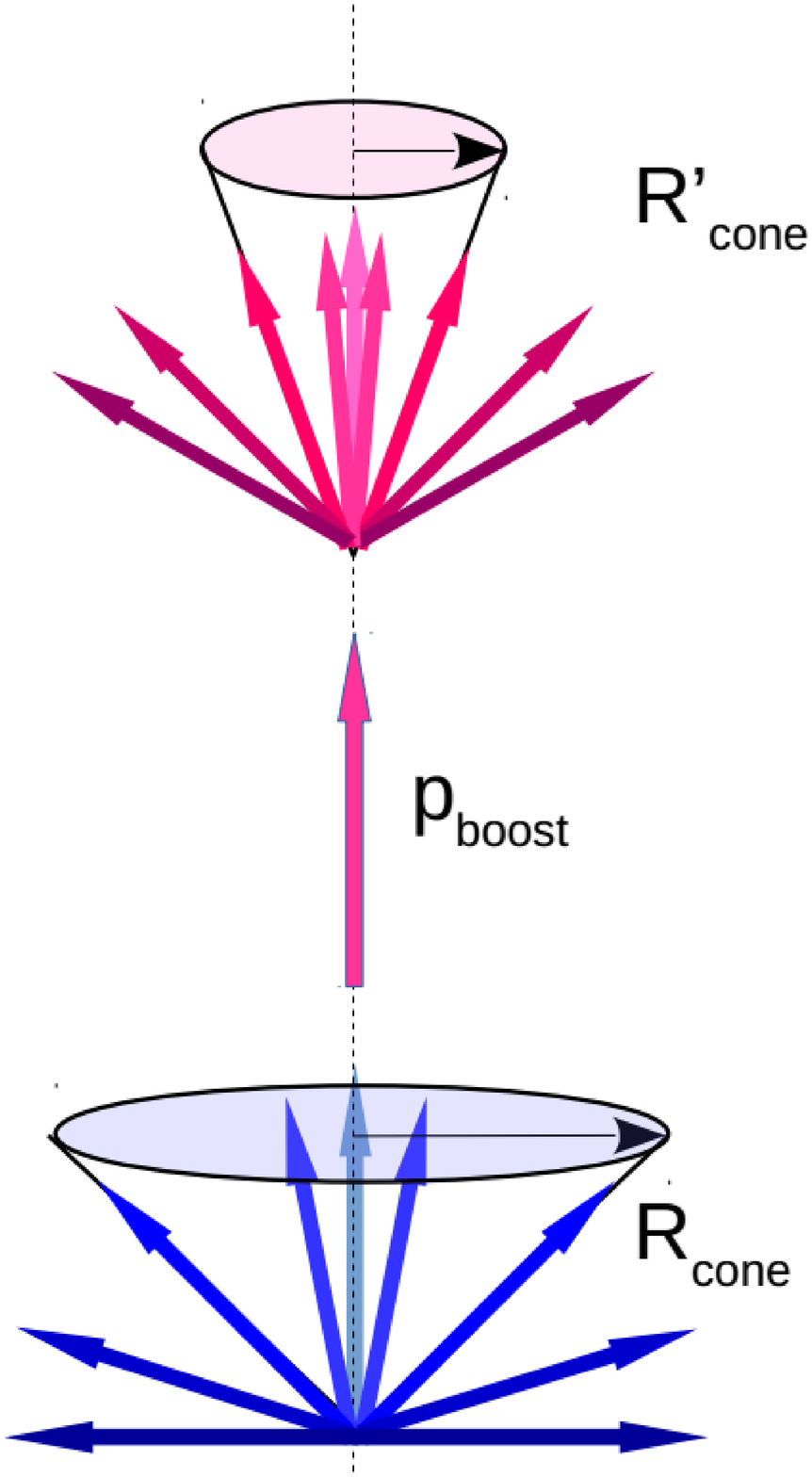}%
\hspace{0.06\linewidth}
\includegraphics[width=0.33\linewidth]{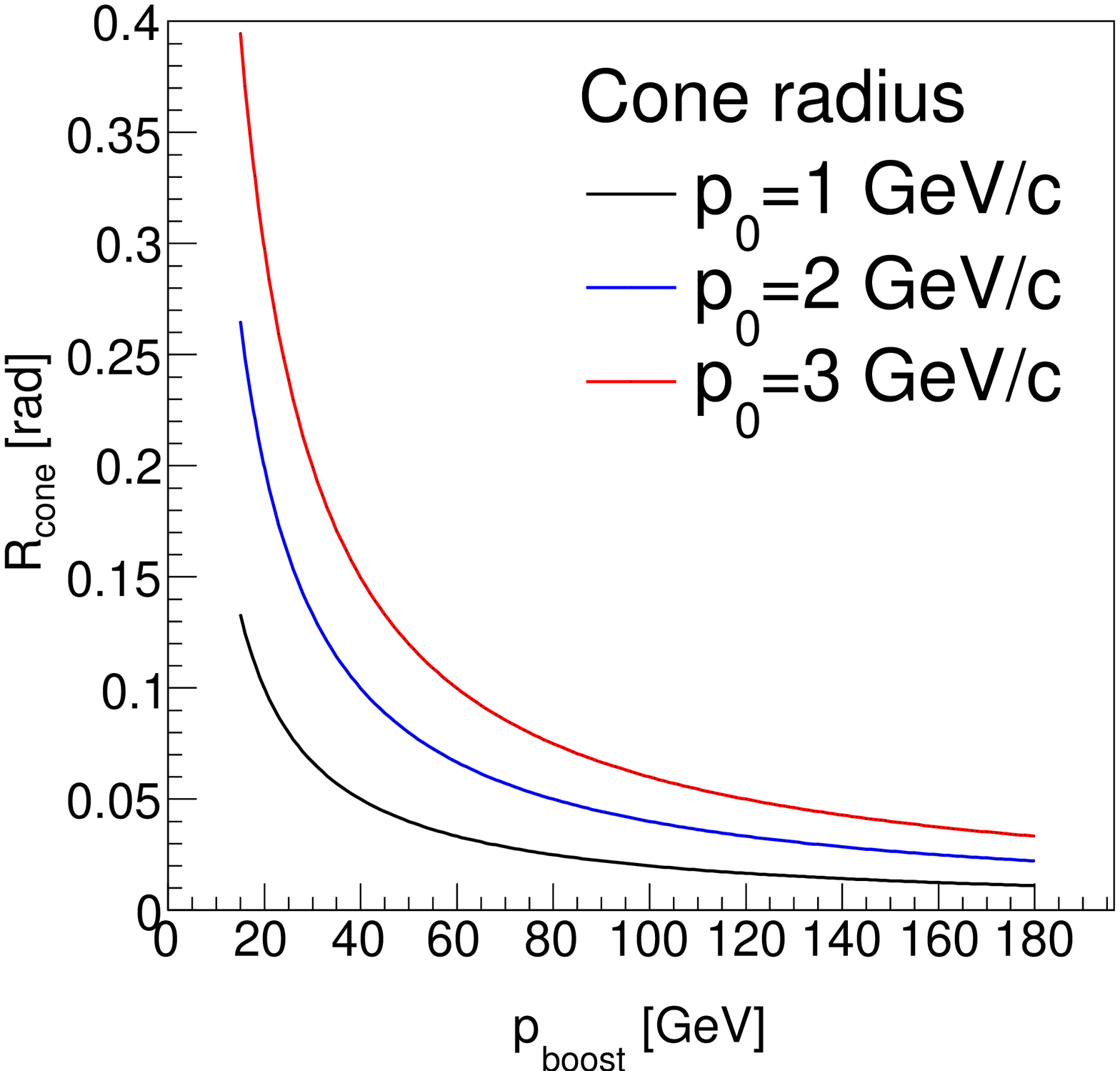}%
\caption{\label{boostedcone}
{\it Left: }Cartoon illustrating the narrowing of a jet cone by Lorentz-boost. {\it Right: }Cone radii according to the simplistic boosted-cone model for different $p_0$ assumptions (see text for details).}
\end{figure}

Jets are more collimated with increasing transverse momentum. In a simple picture this can be linked to Lorentz-boost, ie. the momentum of the initiating parton in the laboratory system.
The \pTjet{}-dependent evolution of $R_\mathrm{fix}$ may also be explained by the Lorentz-boost that high-\pT jets undergo (see the illustration in the left panel of Fig.~\ref{boostedcone}). In order to gain an effect-level understanding, we use a simplistic model. We consider particles radiating from a point in a plane with momenta of equal absolute value $p_0$. We boost these particles along the axis perpendicular to their plane, with a certain momentum $p_\mathrm{boost}$. The resulting particles will form a cone around the boost axis in the lab system, representing our "jet". In the right panel of Fig.~\ref{boostedcone} we see that the resulting size of the "jet", $R_\mathrm{cone}$, depends on $p_\mathrm{boost}$ in a qualitatively similar manner to how the intersection radius $R_\mathrm{fix}$ depends on \pTjet. This attests to the assumption that the universal behavior can, at least partially, be understood by the narrowing by Lorentz-boost of high-\pT{} jets. However the angular cut-off  that limits the jet sizes of low momenta in pQCD-based models is not implemented in our simple toy model, allowing $R_\mathrm{cone}$ to blow up at low $p_\mathrm{boost}$ values. Also, one cannot expect real jets to go below a certain size because after certain point the clustering algorithms will be driven by the presence of UE. This may explain the apparent convergence of the $R_\mathrm{fix}$ curves to a finite value at high \pT. As mentioned before, $R_\mathrm{fix}$ at high-\pT is also influenced by the choice of $\delta r$.
A particularly interesting question is whether $R_\mathrm{fix}$ can be generalized to the larger and more complex systems produced in heavy-ion collisions. To see that, one would need to do simulations in heavy-ion collisions and verify the outcome with data. In  case $R_\mathrm{fix}$ is representative of the jet size in heavy-ion collisions, it would provide a handy observable for the exploration of medium modification of jets.

\begin{figure}[!h]
\includegraphics[width=0.33\linewidth]{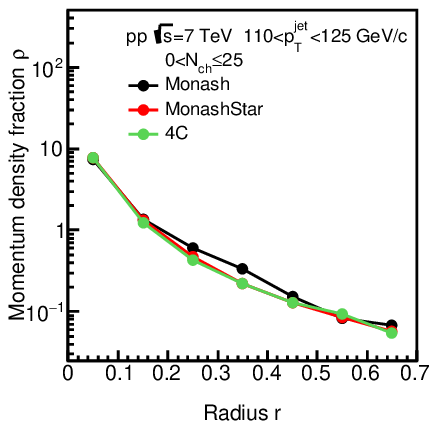}%
\includegraphics[width=0.33\linewidth]{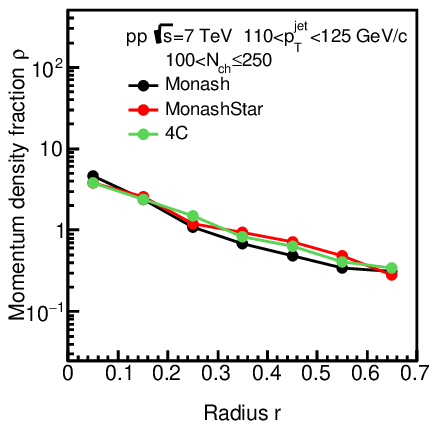}%
\includegraphics[width=0.33\linewidth]{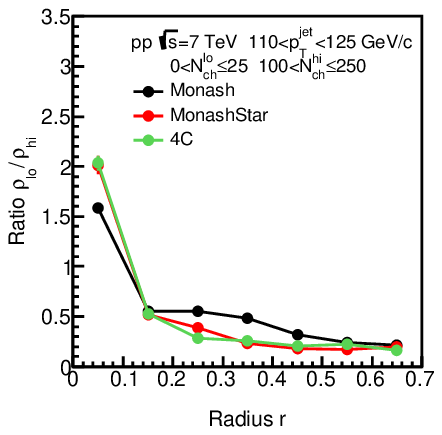}%
\caption{\label{fig:rhoTunes}
Differential jet structure compared for several PYTHIA tunes, for $110\ \GeV/c<\pTjet<125\ \GeV/c $. Jet $\rho(r)$ from events with a lower multiplicity of $\Nch\leq 25$ {\it (left)} are compared to jets from events with a higher multiplicity of $100<\Nch\leq 250$ {\it (center)}. The ratio of the two is shown in the {\it right} panel.
}
\end{figure}

We plotted in Fig.~\ref{fig:rhoTunes} the differential jet structure for various PYTHIA tunes in a particular \pTjet window to compare them in the low and high multiplicity regions. 
In the right panel of Fig.~\ref{fig:rhoTunes} we take the differential jet shapes for the above mentioned low and high multiplicity classes and divide them with each other to highlight the differences for the different tunes. 
As expected, jets from low-multiplicity events have a more steeply falling momentum density distribution than the ones from high-multiplicity events, which is also reflected in a falling ratio. However, there are also certain significant differences between the selected tunes that are beyond this trivial effect.

\begin{figure}[!h]
\includegraphics[width=0.33\linewidth]{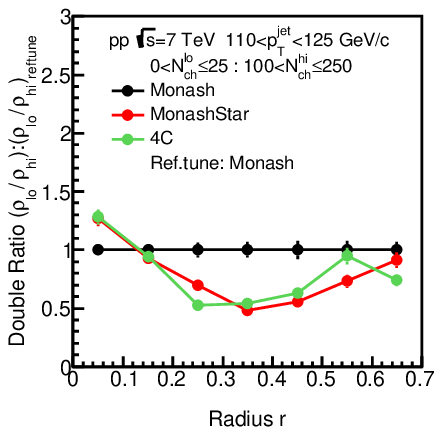}%
\includegraphics[width=0.33\linewidth]{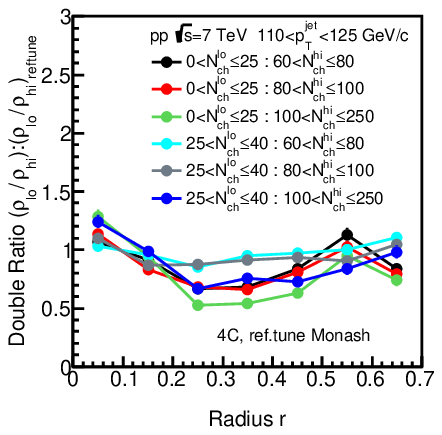}%
\caption{\label{rhoratio4} 
{\it Left:} The double ratio with Monash as the reference tune for the same multiplicity classes. {\it Right:} The double ratio shown for different selections of high and low multiplicity classes (see text in figure).}
\end{figure}

To highlight the differences between the jet structures from different tunes we compute the double ratio,
\begin{equation}
DR(r) = \frac{\rho_\mathrm{low}/\rho_\mathrm{high}}{(\rho_\mathrm{low}/\rho_\mathrm{high})_\mathrm{ref.tune}} \,
\end{equation}
where we divide the former ratio of the high and low multiplicity classes with the very same ratio calculated for the Monash tune. After the trivial effect is gone, a rather sizable effect in the order of a factor of 2 can be seen for both the 4C and the Monash* tunes, w.r.t. the Monash as the reference tune. 
The right panel of Fig.~\ref{rhoratio4} shows the same calculations for the 4C tune, for several different choices of high and low-multiplicity class pairs. In this selected \pTjet range all show similar structures, and generally the effect is larger when the separation in multiplicity is larger. It is very important to note that these curves are derived from statistically independent samples, hence cannot be explained by fluctuations.
Since on Figs~\ref{fig:rhoTunes} and \ref{rhoratio4} we calculate ratios of binned data without a bin center correction, we tested its possible effect by decreasing the bin size from $\delta r = 0.1$ to $\delta r = 0.05$. We did not find any difference beyond statistical uncertainties.

\begin{figure}[!h]
\includegraphics[width=0.33\linewidth]{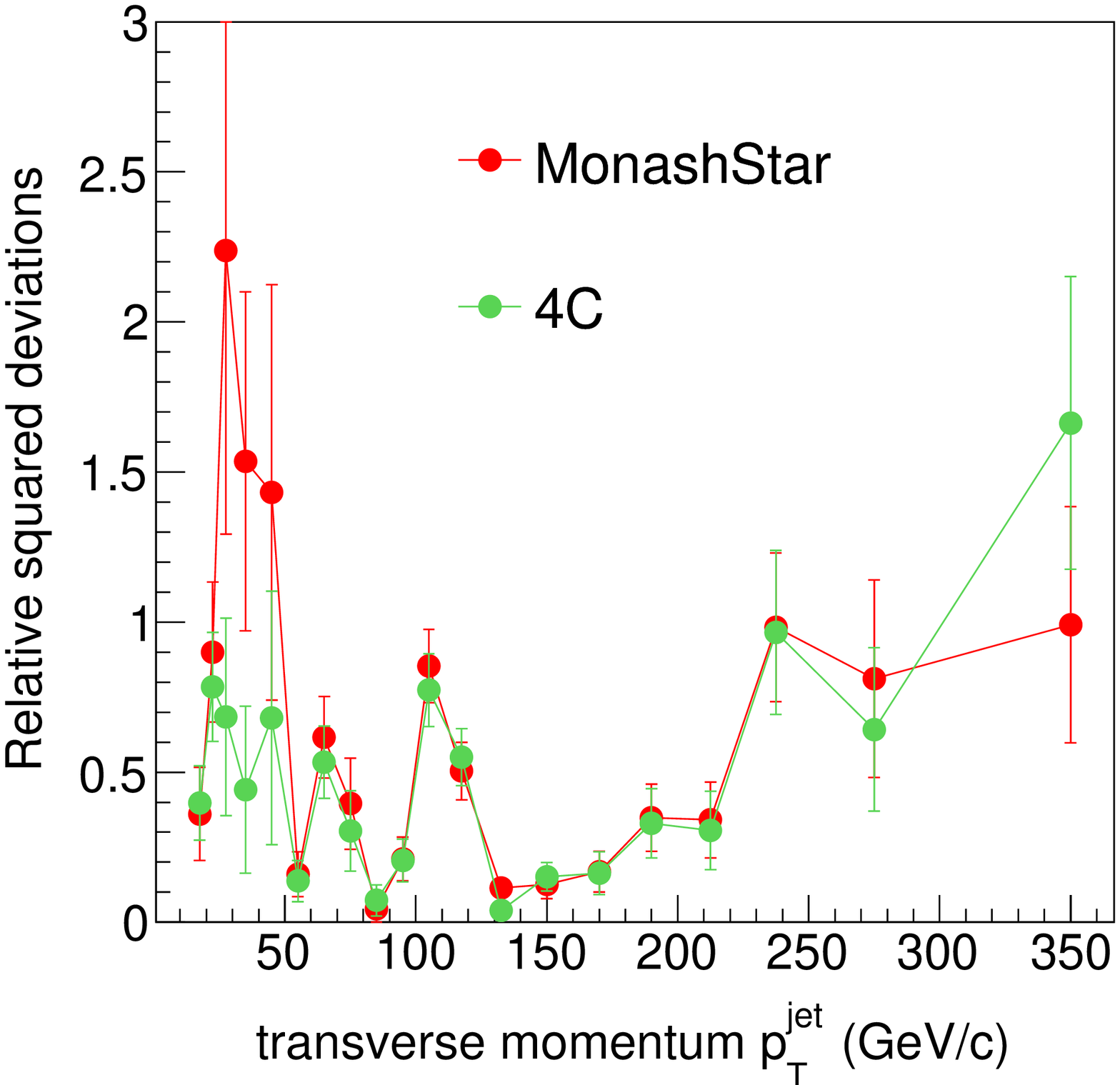}%
\includegraphics[width=0.33\linewidth]{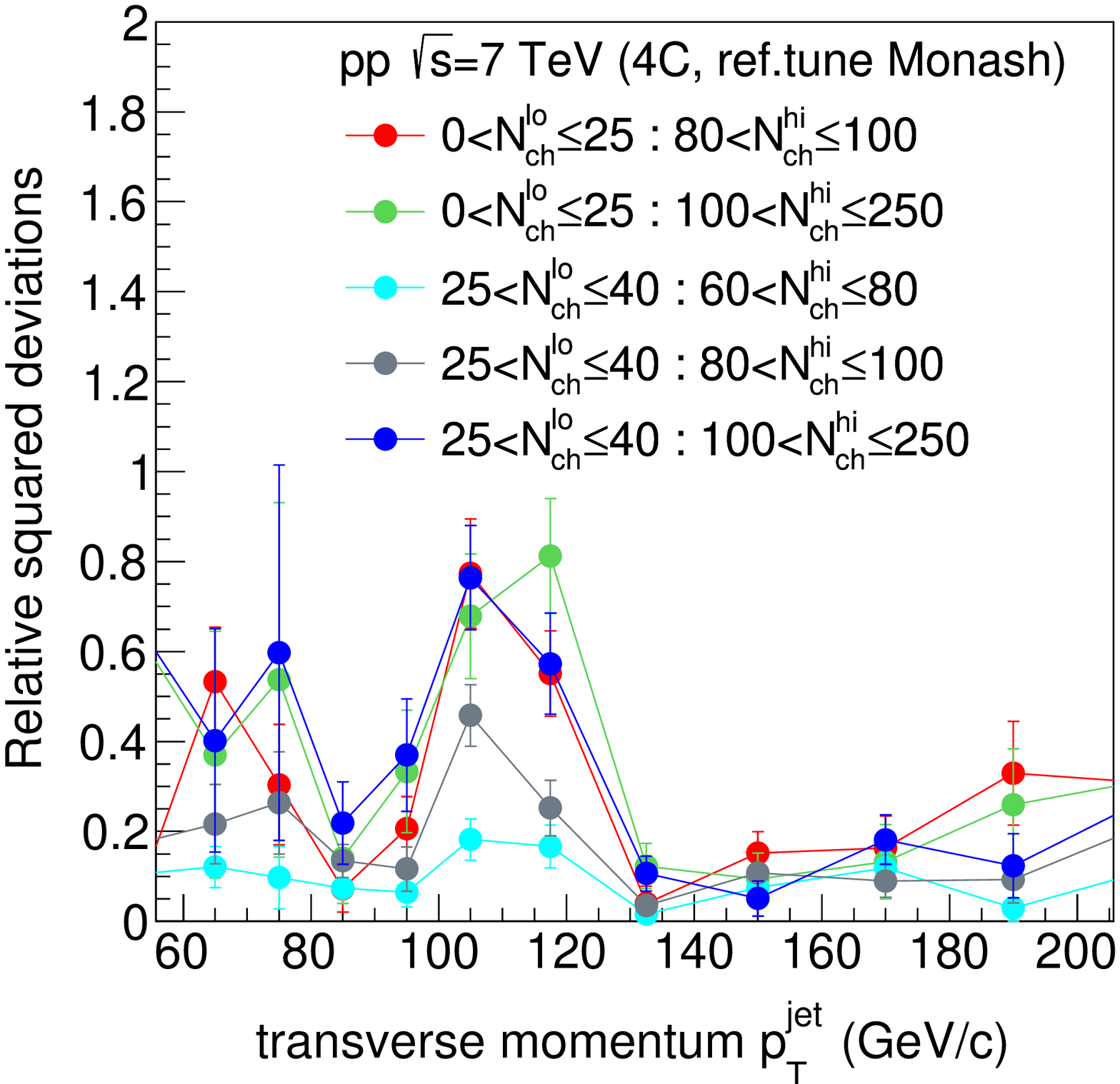}%
\caption{\label{diffscan}
{\it Left:} Square sum of the bins in the $DR(r)-1$ diagram for the 4C (green) and Monash* (red) tunes w.r.t. the Monash reference tune, depending on the \pTjet, for $0<\Nch\leq 25$ as low-multiplicity, and $80<\Nch\leq 100$ as high-multiplicity selections. 
{\it Right:} Square sum of the bins in the $DR(r)-1$ diagram for the 4C tunes w.r.t. the Monash reference tune, depending on the \pTjet, for various low- and high-multiplicity selections, as listed in the legend. (The \pTjet range is restricted to omit parts with large fluctuations.)
}
\end{figure}

In order to understand the dependence of the effect on \pTjet, one might wish to describe the deviations for each \pTjet value with a single number. Therefore we compute the squared sum of the bin-by-bin deviations of the double ratio from the Monash tune, i.e. 
\begin{equation}
RSD=\sqrt{\sum_{0<r_i<R}{\left(DR(r_i)-1\right)^2}}
\end{equation}
at a given \pTjet. In Fig.~\ref{diffscan} we show the results for different tunes as well as for different selections of multiplicity class pairs. Again we see a rather parallel behavior of the 4C and Monash* tunes (or, in other words, the Monash tune is the one that deviates from these two). The behavior vs. \pTjet is non-trivial with several minima and maxima, and is not easily explained without taking into account peculiar details of each tune. However, one sees again a very strong correlation between curves of different multiplicity selections calculated independently from each other, and that the amplitude strongly depends on the separation between the low- and high-multiplicity classes. Thus we can conclude that the multiplicity-dependent analysis of jet structures in a wide \pTjet range has the potential of evaluating the goodness of tunes that otherwise preform equally well in several tests.

\begin{figure}[!h]
\includegraphics[width=0.33\linewidth]{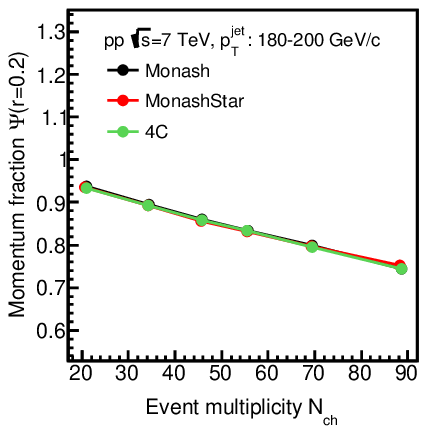}%
\includegraphics[width=0.33\linewidth]{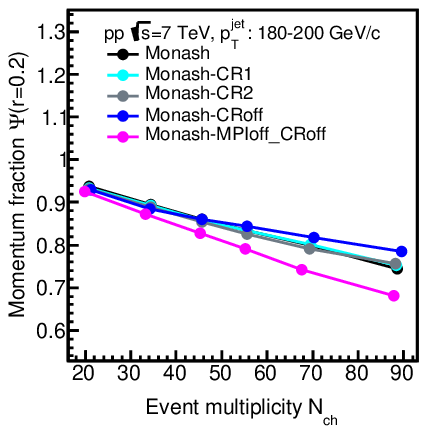}%
\caption{\label{psivsnch}
Evolution of the integral jet structure $\psi(r=0.2)$ with event multiplicity \Nch, at $180\ \GeV/c<\pTjet<200\ \GeV/c$, compared for several PYTHIA tunes {\it (left)} and for the different settings {\it (right)} as explained in the text. The points are placed according to the weight of the distribution in each multiplicity class.}
\end{figure}

Lastly, in Fig.~\ref{psivsnch}, the integral jet shape is plotted in
terms of the multiplicity at high momenta, $180\ \GeV/c<\pTjet<200\ \GeV/c$. In the {\it left panel}, where tunes are
compared, there is no observable effect in the integral structure between the tunes Monash, Monash* and 4C. 
We present the effects of different MPI and CR settings on the integral jet structure in the {\it right panel} of Fig.~\ref{psivsnch}.
Different color reconnection schemes do not lead to significant differences, but there is a slight deviation at high \Nch values when
color reconnection is turned off. However, the lack of MPI causes a significant difference within the
same multiplicity class, that grows approximately linearly with \Nch, which suggests that the MPI has a strong influence on the jet
structure, especially at high \Nch values. It is to be noted that the effect is less significant in case of lower \pTjet windows and in case
of larger $r$ values. At lower multiplicties, MPI and CR cause little difference in the integrated jet shape. That the $\psi(r)$ values at high \Nch are lower
in the case the MPI is turned off, means the jets are more concentrated in a narrow cone. This can be understood by a higher relative fraction
of soft tracks coming from the UE in case when there is no MPI, compared to the MPI case with the same multiplicity where there is a more relevant contribution from tracks that come from the jet itself. 
Note that the points in Fig.~\ref{psivsnch} are not at the bin centers, but they are placed to represent the weight of the \Nch
distribution in a given bin, to eliminate the possible bias stemming from the different \Nch distributions within multiplicity classes.

Understanding the observed dependence of the integrated jet structure on the multiplicity needs further analysis supported by experimental data. The above observation, if compared to real data, may provide a control over the extent of MPI effects. Further studies are needed to identify MPI/CR effects and separate them from the UE, also using other observables that are less sensitive to the UE.

\section{Summary}
\label{sec:summary}

\red{We performed a novel jet shape analysis in $\sqs=7$~\TeV pp collisions to explore the multiplicity and 
\pTjet-dependence of differential and integrated jet structure observables. 
We used several models} implemented in the PYTHIA 8.226 event generator. We demonstrated that the simulations describe CMS data, and we gave predictions for \red{the} jet structure observables in several multiplicity classes, over a wide momentum range. We found that there is a given radius $R_\mathrm{fix}$ where jet momentum density is independent of multiplicity. This radius is insensitive to the choice of simulation settings \red{(choice of tune, presence and modelling of MPI and CR)} within the investigated model class and even of jet clustering algorithms, and its \pT{}-dependence qualitatively follows a Lorentz boost curve. \red{These observations suggest that $R_\mathrm{fix}$ is an inherent property of jets that is characteristic to the spatial development of the parton shower at a given momentum.}

We compared the multiplicity dependence of jet structure variables for three popular PYTHIA tunes as well as different MPI and CR models in several \pT bins. We found that the evolution of \red{the} differential jet structure $\rho(r)$ with multiplicity significantly differ in several \pTjet ranges for the Monash, Monash* and 4C tunes. The shape of the difference is nontrivial in \pTjet, but persistent through all tested choices of multiplicity selections. With this we demonstrated that the multiplicity-dependent analysis of jet momentum profiles can differentiate among otherwise well-established models. \red{This lack of understanding may have grave consequences on studies based on classification by jet properties. Our} observation highlights the need of extending multiplicity-dependent jet structure measurements such as in Ref.~\cite{Chatrchyan:2013ala} to higher \pTjet regimes.

We also see that the integrated jet structure variable $\psi(r=0.2)$ shows a rather different \Nch{}-dependence when MPI are turned off. This attests to the important role of multiple-parton interactions in higher multiplicity events and the need for their detailed understanding in order to develop accurate models in jet physics.

\acknowledgements
The Authors are thankful for the enlightening discussions they had with Yaxian Mao and Jana Biel\v{c}\'\i{}kov\'a. This work has been supported by the Hungarian National Research Fund (OTKA) grant K120660 and THOR COST action 15213. We acknowledge the support of the Wigner GPU laboratory and the Wigner Datacenter. Author RV thanks for the support of the J\'anos Bolyai fellowship of the Hungarian Academy of Sciences.

\bibliographystyle{elsarticle-num}
%


\end{document}